\documentclass[twocolumn,english,aps,prb,superscriptaddress,groupedaddress]{revtex4}
\usepackage{amsmath}
\usepackage{amssymb}
\usepackage{graphicx}

\makeatletter

\usepackage{color}

\makeatother

\begin{document}

\title{Majorana fermions in topological insulator nanowires: from single
superconducting nanowires to Josephson junctions}

\author{Guang-Yao~Huang}
\affiliation{Beijing Key Laboratory of Quantum Devices, Key Laboratory for the Physics and
Chemistry of Nanodevices, and Department of Electronics, Peking University, Beijing 100871, China}

\author{H.~Q.~Xu}
\email[Corresponding author. ]{hqxu@pku.edu.cn; hongqi.xu@ftf.lth.se}
\affiliation{Beijing Key Laboratory of Quantum Devices, Key Laboratory for the Physics and
Chemistry of Nanodevices, and Department of Electronics, Peking University, Beijing 100871, China}
\affiliation{Division of Solid State Physics, Lund University, Box 118, S-221 00 Lund, Sweden}

\date{\today}
\begin{abstract}
Signatures of Majorana fermion bound states in one-dimensional topological insulator (TI)
nanowires with proximity effect induced superconductivity are studied. The phase diagram and energy spectra are calculated for single TI nanowires and it is shown that the nanowires can be in the topological invariant phases of winding numbers  $W=0, \pm 1$, and $\pm 2$ corresponding to the cases with zero, one and two pairs of Majorana fermions in the single TI nanowires.
It is also shown that the topological winding numbers, i.e., the numbers of pairs of Majorana fermions in the TI nanowires can be extracted from the transport measurements of a Josephson junction device made from two TI nanowires, while the sign in the winding numbers can be extracted using a superconducting
quantum interference device (SQUID) setup.
\end{abstract}
\maketitle

\section{Introduction}

A pioneered theoretical proposal for realizing
Majorana fermions (MFs) in a topological insulator (TI) \cite{prl100.096407}
raised soon after the concept of TIs was introduced \cite{prl95.146802}
and was studied extensively \cite{rmp82.3045,rmp83.1057,sst27.124003,rpp75.076501,arcmp4.113,rmp87.137} in recent years. Yet signatures of MFs in two- or three-dimensional
TIs still need to be experimentally confirmed, theoretical study of MFs in one-dimensional
(1D) TIs has been greatly inspired by recent progresses
in experimental search for MFs in semiconductor nanowires (NWs).\cite{nl12.6414,science336.1003,nphys8.795,nphys8.887,prl110.126406,science346.602,sr4.7261,nature531.206,arxiv1603.04069,science354.1557}

The topological invariant of MF bound states (the number
of MF pairs) in a 1D system could be classified as $\mathbb{Z}$
or $\mathbb{Z}_{2}$ according to the symmetries of the system.\cite{rmp82.3045,njp12.065010}
For a well-known example of a spinless p-wave superconductor NW 
or a single-band Rashba spin-orbit interaction (SOI)
NW in the proximity of an s-wave superconductor,\cite{pu44.131,prl105.077001,prl105.177002}
the topological classification is $\mathbb{Z}_{2}$
with a single pair of MFs in the topological phase. We are interested
in the 1D systems with more than one pairs of MFs. Multiple pairs of MFs
are found theoretically in systems other than 1D TIs,
such as multiband NWs \cite{prb83.155429,prb84.144522,prl106.127001,prb91.094505,prb91.121413} and
a quantum Ising chain with long range interaction and time reversal
symmetry.\cite{prb85.035110} Of a special type are
the Kramers pairs of MFs, which are two pairs of MFs with each
pair being the time reversal counterpart of the other pair, found in
coupled NWs,\cite{prl111.116402,prl112.126402,prb91.121413,prb92.014514}
time-reversal-invariant topological superconductors,\cite{prl108.036803,prl111.056402,prx4.021018} interacting bilayer Rashba systems,\cite{prl108.147003} $\pi$ Josephson
junctions  made of TIs \cite{prb90.155447,prl115.237001}, and NWs in the proximity of an unconventional superconductor.\cite{prb86.184516,prl110.117002}

Here, we present the topological phase diagram of TI NWs and demonstrate that it is possible to generate both single and multiple pairs of MFs in a TI NW. Our approach starts from
the bulk TI model and employs a dimensional reduction scheme to model a TI nanowire, following
the procedure employed in studying the Rashba SOI nanowires in Refs.~\onlinecite{prl105.077001} and \onlinecite{prl105.177002}. The \mbox{Pfaffian} approach \cite{pu44.131,prb88.075419}
is not sufficient due to the fact that the energy bandgap closing happens not only
in specific points in the Brillouin zone, e.g., the zone center with momentum
$q=0$ and the zone boundary with $q=\pi$, but also in between. We follow Tewari and Sau \cite{prl109.150408} and, after revealing
the chiral symmetry, we obtain all the information of the phase diagram of TI NWs
including the phase boundary and winding number in each phase. We also calculate the energy
spectrum of a TI NW Josephson junction structure and show that a significant difference between one and two pairs of MFs and the sign ``$\pm$'' in the front of the topological winding number $\mathbb{Z}$ can be distinguished:
the number of MF pairs $\mathbb{Z}$ in a TI NW can be mapped to the number of pairs of
accompanied subgap bound states
in the Josephson
junction \cite{prb87.104509} and the sign in $-\mathbb{Z}$ represents the extra $\pi$ phase difference
of the MFs when comparing with the MFs in the $\mathbb{Z}$ topological phase, which
could lead to a topological $\pi$ Josephson junction \cite{prb87.100506} and
could be observed through tunnel spectroscopy measurements
of a superconducting quantum interference device (SQUID) structure.

\section{Formalism}

In difference from the theoretical methods of Refs.~\onlinecite{prl105.136403} and \onlinecite{prb84.201105}, our formalism starts from the three-dimensional bulk TI model and employs a dimensional reduction scheme.
The three-dimensional
bulk TI Hamiltonian is \cite{nphys5.438,prb82.045122,jap110.093714}
\begin{equation}
H_{0}=\varepsilon(\boldsymbol{k})+M(\boldsymbol{k})\Gamma_{5}+B_{0}k_{z}\Gamma_{3}+A_{0}(k_{x}\Gamma_{1}+k_{y}\Gamma_{2}),
\end{equation}
where $\varepsilon(\boldsymbol{k})=C_{0}+C_{1}k_{z}^{2}+C_{2}k_{i}^{2}$,
$M(\boldsymbol{k})=M_{0}+M_{1}k_{z}^{2}+M_{2}k_{i}^{2}$, and $k_{i}^{2}=k_{x}^{2}+k_{y}^{2}$.
The Dirac $\Gamma$ matrices are $\Gamma_{1}=\sigma_{x}\otimes\tau_{x}$,
$\Gamma_{2}=\sigma_{y}\otimes\tau_{x}$, $\Gamma_{3}=\sigma_{z}\otimes\tau_{x}$,
$\Gamma_{4}=\sigma_{0}\otimes\tau_{y}$, and $\Gamma_{5}=\sigma_{0}\otimes\tau_{z}$,
with $\sigma_{i}$ and $\tau_{i}$ (here $i=x,y,z$) being the Pauli matrices acting
on the spin and the parity space and satisfying the Clifford algebra
$\{\Gamma_{i},\Gamma_{j}\}=2\delta_{i,j}$. The Hamiltonian is presented in the basis of $\{\psi_{1\uparrow},\psi_{2\uparrow},\psi_{1\downarrow},\psi_{2\downarrow}\}^{T}$,
where $\psi_{\alpha,\sigma}\; (\alpha =1 \; \mbox{or} \; 2 \; \mbox{and} \; \sigma=\uparrow \; \mbox{or} \; \downarrow)$ is a
$p$-like orbital $\alpha$ with spin $\sigma$, and is invariant under the time
reversal operation $\mathcal{T}=i\sigma_{y}\otimes\tau_{0}\mathcal{K}$, where
$\mathcal{K}$ is the complex conjugate operator.
By dimensional reduction, the 1D form of the Hamiltonian (defined along the $x$ direction) can be written as
\begin{equation}
\mathcal{H}_{0}=C_{0}+C_{2}k_{x}^{2}+(M_{0}+M_{2}k_{x}^{2})\Gamma_{5}+A_{0}k_{x}\Gamma_{1}.
\end{equation}
In the presence of an external magnetic field along the $z$ direction, the Zeeman term is
\begin{eqnarray}
\mathcal{H}_{z} & = & \mathrm{diag}(V_{z1},V_{z2},-V_{z1},-V_{z2})\nonumber \\
 & = & V_{z}\mathrm{diag}(1,g,-1,-g),
\end{eqnarray}
where $g=V_{z2}/V_{z1}$ is the ratio of the Zeeman energies of the two $p$-like orbitals.
Considering the proximity effect by an s-wave superconductor, the term $H_{sc}$ which describes the effect of pairing is
\begin{eqnarray}
H_{sc}= \int & dx & \left[ \Delta_1 e^{-i\theta} (\psi_{1\uparrow}^{\dagger}\psi_{1\downarrow}^{\dagger}+\psi_{2\uparrow}^{\dagger}\psi_{2\downarrow}^{\dagger}) \right. \nonumber \\
& + & \left. \Delta_2 e^{-i\theta} (\psi_{1\uparrow}^{\dagger}\psi_{2\downarrow}^{\dagger}+\psi_{2\uparrow}^{\dagger}\psi_{1\downarrow}^{\dagger})+h.c.\right],
\end{eqnarray}
where $\Delta_1$ is the amplitude of pairing potential between the same orbitals and $\Delta_2$ between different orbitals in the Bardeen-Cooper-Schrieffer (BCS) scenario, and $\theta$ is the phase in the paring potentials.

The Bogoliubov-de Gennes (BdG) Hamiltonian of a 1D TI NW is obtained by representing the total Hamiltonian, $\mathcal{H}_{0}+\mathcal{H}_{z}+\mathcal{H}_{sc}$, in the Nambu basis $\Psi=\left[\begin{array}{c}
\psi\\
\mathcal{T}\psi
\end{array}\right]$
as
\begin{equation}
\mathcal{H}_{BdG}=\left[\begin{array}{cc}
\mathcal{H}_{0}+\mathcal{H}_{z} & \Delta_0 e^{-i\theta}\\
\Delta_0 e^{i\theta} & -(\mathcal{H}_{0}+\mathcal{H}_{z})^{\mathcal{T}}
\end{array}\right],
\end{equation}
with
\begin{equation}
\Delta_0=\Delta_1\Gamma_{0}-i\Delta_2\Gamma_{4}\Gamma_{5},
\end{equation}
where $\Gamma_0$ is the identity matrix. Introducing another set of Pauli matrices $\rho_{i}(i=x,y,z)$ acting
on the particle-hole space, the BdG Hamiltonian can be rewritten as
\begin{equation}
\mathcal{H}_{BdG}=\mathcal{H}_{0}\rho_{z}+\mathcal{H}_{z}+\Delta_{0}(\rho_{x}\cos\theta-\rho_{y}\sin\theta).
\end{equation}
The system can be solved numerically by discretizing $\mathcal{H}_{BdG}$ into a lattice model,
\begin{eqnarray}
(\mathcal{H}_{BdG})_{mn} & = & [(2t+2\eta\Gamma_{5}+\zeta\Gamma_{5}-\mu)\rho_{z}\nonumber \\
 & + & \mathcal{H}_{z}+\Delta_{0}(\rho_{x}\cos\theta-\rho_{y}\sin\theta)]\delta_{m,n}\nonumber \\
 & - & (t+\eta\Gamma_{5}+i\alpha\Gamma_{1})\rho_{z}\delta_{m+1,n}\nonumber \\
 & + & (-t+\eta\Gamma_{5}+i\alpha\Gamma_{1})\rho_{z}\delta_{m-1,n},
\end{eqnarray}
where parameters $t=C_{2}/a^{2}$, $\mu=-C_{0}$, $\zeta=M_{0}$,
$\eta=M_{2}/a^{2}$, and $\alpha=A_{0}/2a$.

\begin{figure}
\centering\includegraphics[scale=0.8]{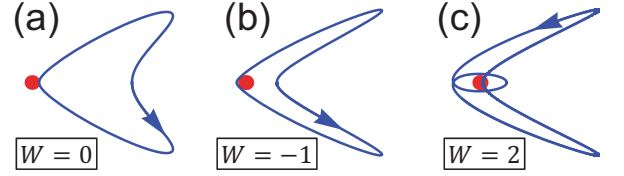}\caption{\label{windingCurve}(a) to (c) Sketches for winding around a zero (marked by a red dot) of $z(q)$ in complex plane for three examples with the
winding numbers of  $W= 0$, $-1$ and $2$.}
\end{figure}

For an infinitely long TI NW, the above Hamiltonian can be written as
\begin{equation}
\mathcal{H}_{BdG}(q)=\tilde{\mathcal{H}}{}_{0}(q)\rho_{z}+\mathcal{H}_{z}+\Delta_{0}\rho_{x}
\end{equation}
where the phase $\theta$ is removed
by a global gauge transformation which is valid when considering a single TI NW segment
{[}but will be reinstalled later in the study of a TI NW Josephson junction structure{]} and $\tilde{\mathcal{H}}{}_{0}(q)=2t-\mu+2\eta\Gamma_{5}+\zeta\Gamma_{5}-2(t+\eta\Gamma_{5})\cos q+2\alpha\Gamma_{1}\sin q$, and $q$ the dimensionless momentum.
The fact that $\mathcal{H}_{BdG}(q)$ is real allows us to make a unitary
transformation $U$ in particle-hole space, $\mathcal{H}'_{BdG}(q)=U\mathcal{H}_{BdG}(q)U^{\dagger}$, to obtain a Hamiltonian of block off-diagonal form,
\begin{equation}
\label{hBdGPrime}
\mathcal{H}'_{BdG}(q)=\left[\begin{array}{cc}
0 & A(q)\\
A^{*}(q) & 0
\end{array}\right],
\end{equation}
where $U=[e^{-i\rho_{z}\frac{\pi}{4}}e^{-i\rho_{x}\frac{\pi}{4}}\otimes \Gamma_0][\Gamma_0\oplus(-i\Gamma_{2}\Gamma_{3})]$,
$A(q)=\tilde{\mathcal{H}}{}_{0}(q)+\mathcal{H}_{z}-i\tilde{\Delta}_{0}$
with $\tilde{\Delta}_{0}=-i\Delta_1\Gamma_{2}\Gamma_{3}+\Delta_2\Gamma_{1}$.
After defining a complex number
\begin{equation}
z(q)=\mathrm{Det}(A(q)),
\end{equation}
the topological invariant winding number $W$ can be calculated through
\begin{equation}
W=\frac{1}{2\pi i}\oint\frac{dz(q)}{z(q)}.
\end{equation}
Thus, $W$ can be simply evaluated by counting how many circles surrounding a zero
point of $z(q)$ in complex plane that have taken place over one period, see Fig.~\ref{windingCurve} for three typical examples.

\section{Phase diagrams and energy spectra of single superconducting TI Nanowires}


\begin{figure}
\centering\includegraphics[scale=0.9]{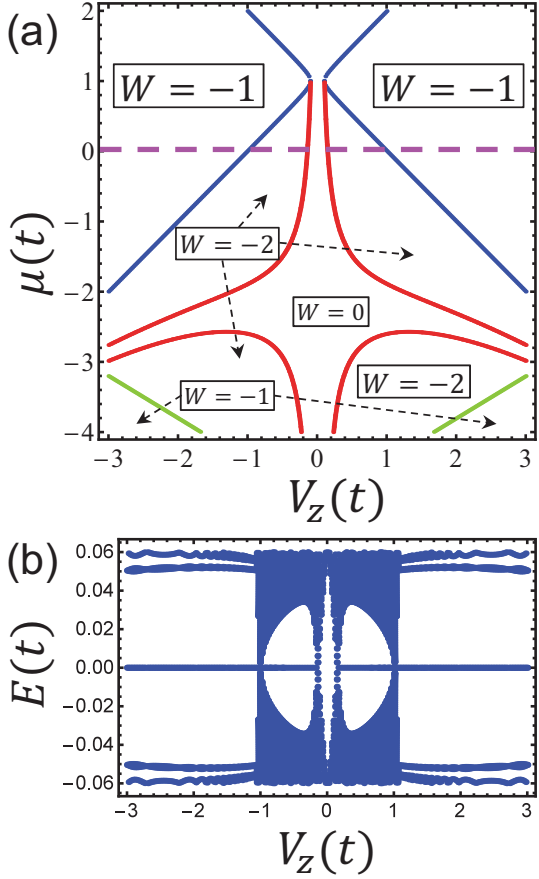}
\caption{\label{phaseAndSpectraCase1}
(a) Phase diagram of a 1D superconducting TI in the chemical potential-Zeeman energy ($\mu-V_{z}$) plane in the case of  $\Delta_1>\Delta_2$. The 1D superconducting TI is modeled with parameters $\alpha/t=2$, $\zeta/t=5$, $\eta/t=-2$, $\Delta_1/t=0.1$,
$\Delta_2/t=0.05$, and $g=0.6$. The green, blue and red lines mark the phase boundaries defined by gap closing at $q=0$, $\pi$, and a value in between in the Brillouin zone, respectively. The corresponding winding numbers in different phase regions are indicated in the figure.
(b) Energy spectra $E$ vs Zeeman energy $V_{z}$ of a corresponding finite superconducting TI NW at $\mu=0$, corresponding to the dotted line in (a). In the calculations, the finite superconducting TI NW is modeled by a lattice of 800 sites.
}
\end{figure}

The phase diagrams and the energy spectra of 1D superconducting TIs are calculated. We first present and discuss the calculations in the presence of a magnetic field and in the case of $\Delta_1>\Delta_2$. Figure~\ref{phaseAndSpectraCase1}(a) shows the calculated phase diagram for such a TI NW of infinite length. Three kinds of phase boundaries are present in the system: the boundaries defined by (1) gap closing at $q=0$,
(2) gap closing at $q=\pi$, and (3) gap closing when $q$ is at a value between $0$ and $\pi$. The first two kinds
of phase boundaries are due to the particle-hole symmetry, while the third
kind roots from the existence of both $\Delta_1$ and $\Delta_2$.
The phase index $W$ changes by
1 or -1 by crossing the first two kinds of boundaries, but it changes by 2 or -2 by crossing
the third kind of boundaries.
The first two kinds of
boundaries can also be extracted through the Pfaffian approach: $\mathrm{Pf}B(0)=0$
and $\mathrm{Pf}B(\pi)=0$, where $B(q)$ is the antisymmetric matrix
$B(q)=\mathcal{H}_{BdG}(q)(\rho_{y}\otimes\sigma_{y}\otimes\tau_{0})$
and $\mathrm{Pf}$ is the Pfaffian operator.
Some analytic expressions at the first kind of the phase boundaries are available. For example, for $q=0$ and
$g=1$ (other parameters are in units of $t$), the phase boundary is defined by
\begin{equation}
V_{z}^{2}=\Delta_2^{2}+\Delta_1^{2}+\zeta^{2}+\mu^{2}\pm2\sqrt{\Delta_2^{2}(\Delta_1^{2}+\zeta^{2})+\zeta^{2}\mu^{2}}.
\end{equation}
At $\mu=0$, the above equation reads
\begin{equation}
V_{z}^{2}=\Delta_2^{2}+\Delta_1^{2}+\zeta^{2}\pm2\Delta_2\sqrt{\Delta_1^{2}+\zeta^{2}}.
\end{equation}
Figure \ref{phaseAndSpectraCase1}(b) shows the energy spectra of a corresponding finite superconducting TI NW at
the chemical potential $\mu=0$. It is seen that the zero-energy states exist inside the gap in both the $|W|=1$ and the $|W|=2$ phases.


\begin{figure}
\centering\includegraphics[scale=1]{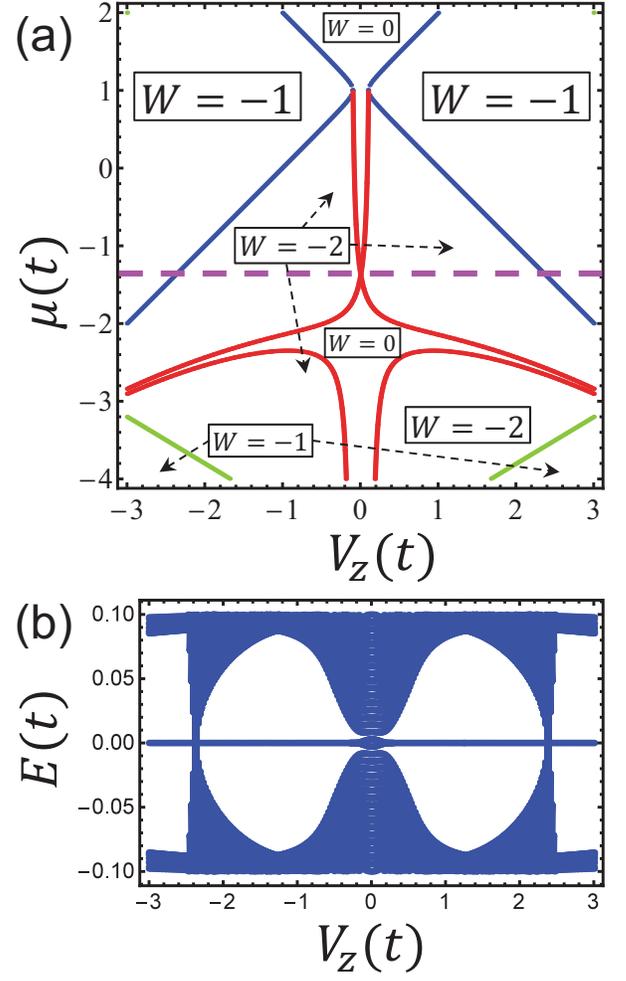}
\caption{\label{phaseAndSpectraCase1d5}
(a) Phase diagram of a 1D superconducting TI in the chemical potential-Zeeman energy ($\mu-V_{z}$) plane in the case of  $\Delta_1=\Delta_2$. Other parameters are the same as in Fig.~\ref{phaseAndSpectraCase1}(a) except for $\Delta_2/t=0.1$. The corresponding winding numbers in different phase regions are indicated in the figure.
(b) Energy spectra $E$ vs Zeeman energy $V_{z}$ of a corresponding finite superconducting TI NW at a value of $\mu\approx -1.373$ corresponding to the dotted line in (a). In the calculations, the finite superconducting TI NW is modeled by a lattice of 1600 sites. Note that at $\mu\approx -1.373$, two phase boundaries that separate the $|W|=2$ and 0 phases are touched at $V_z =0$ as shown in (a).
}
\end{figure}

Figure~\ref{phaseAndSpectraCase1d5}(a) shows the phase diagram of the infinite TI NW in the case of $\Delta_1=\Delta_2$. It is seen that the phase diagram inherits most of the features seen in the case of $\Delta_1>\Delta_2$, such as the three kinds of phase boundaries and the phases with single and multiple pairs of MFs. A significant difference arising from the relative increase of $\Delta_2$ is that the phase boundaries that separate the $|W|=2$ and 0 phases are shifted and two of them are touched to each other at the line of $V_z =0$. Figure~\ref{phaseAndSpectraCase1d5}(b) shows the energy spectra of a corresponding finite superconducting TI NW at the point of $\mu$ for which the two boundaries that separate the $|W|=2$ and 0 phases are touched. It is shown that the region in which the zero-energy states are present extends to almost the entire considered magnetic field range except for the neighborhood of $V_z=0$.


\begin{figure}
\centering\includegraphics[scale=1]{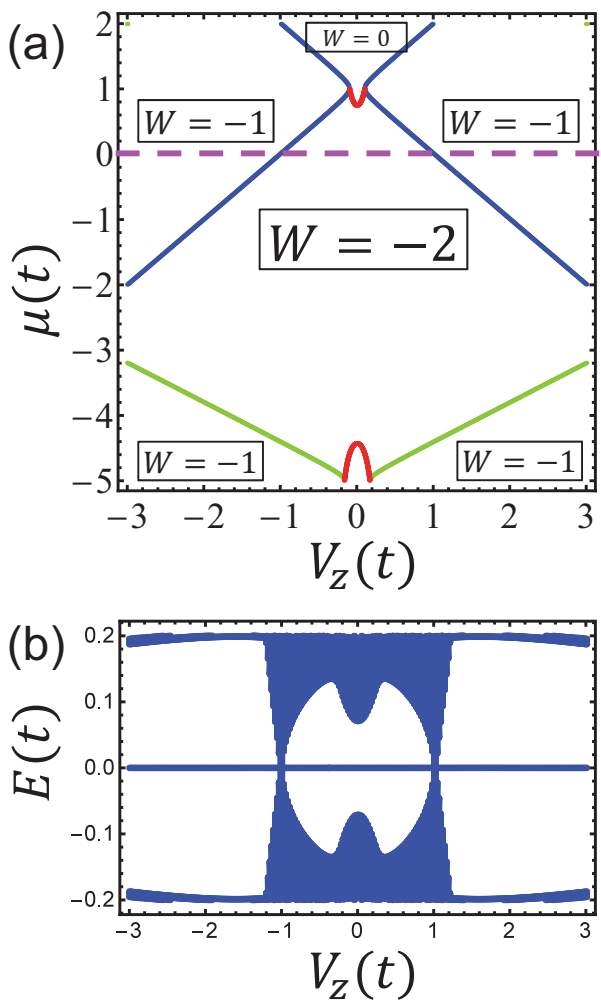}
\caption{\label{phaseAndSpectraCase2}
(a) Phase diagram of a 1D superconducting TI in the chemical potential-Zeeman energy ($\mu-V_{z}$) plane in the case of $\Delta_1<\Delta_2$. Other parameters are the same as in Fig.~\ref{phaseAndSpectraCase1}(a) except for $\Delta_2/t=0.2$. The corresponding winding numbers in different phase regions are indicated in the figure.
(b) Energy spectra $E$ vs Zeeman energy $V_{z}$ of a corresponding finite superconducting TI NW at $\mu=0$, corresponding to the dotted line in (a). The finite superconducting TI NW is modeled by a lattice of 800 sites.
}
\end{figure}

We now consider the 1D superconducting TI NWs in the case of $\Delta_1<\Delta_2$. Figure~\ref{phaseAndSpectraCase2}(a) shows the phase diagram of an infinite superconducting TI NW in this case. It is seen that there still exist the phases with one pair ($|W|=1$) and two pairs ($|W|=2$) of MFs and all the three kinds of phase boundaries are present. However, by comparisons to the phase diagrams shown for the cases of $\Delta_1>\Delta_2$ in Fig.~\ref{phaseAndSpectraCase1}(a) and of $\Delta_1=\Delta_2$ in Fig.~\ref{phaseAndSpectraCase1d5}(a), we see that although the first two kinds of boundaries remain mostly unchanged, the third kind of boundaries have been shifted dramatically, making the $|W|=2$ phase extended to the zero Zeeman energy region. One important consequence of this is that the MFs can emerge at zero magnetic field. Figure~\ref{phaseAndSpectraCase2}(b) shows the energy spectra of a corresponding finite superconducting TI NW along the line of chemical potential $\mu=0$. It is seen that both one pair and two pairs of MFs at zero energy are present inside the gap and the zero-energy MF states appear over the entire range of  considered magnetic fields.

\section{Kramers pairs of MFs}

From the above results and discussion, we can see that the relative strength of $\Delta_1$ and $\Delta_2$ plays an important role in the realization of a 1D superconducting TI system  with the existence of MFs at zero magnetic field. In this section, we will show that the two pairs of MFs in the $|W|=2$ phases at zero magnetic field are Kramers pairs of MFs due to the presence of time reversal symmetry in the system. To be concise, in the following of this section, the Zeeman energy is set to zero ($V_z=0$) and $r$ is defined as the ratio of $\Delta_2$ to $\Delta_1$ ($r=\Delta_2 /\Delta_1$).

\begin{figure}
\centering\includegraphics[scale=1]{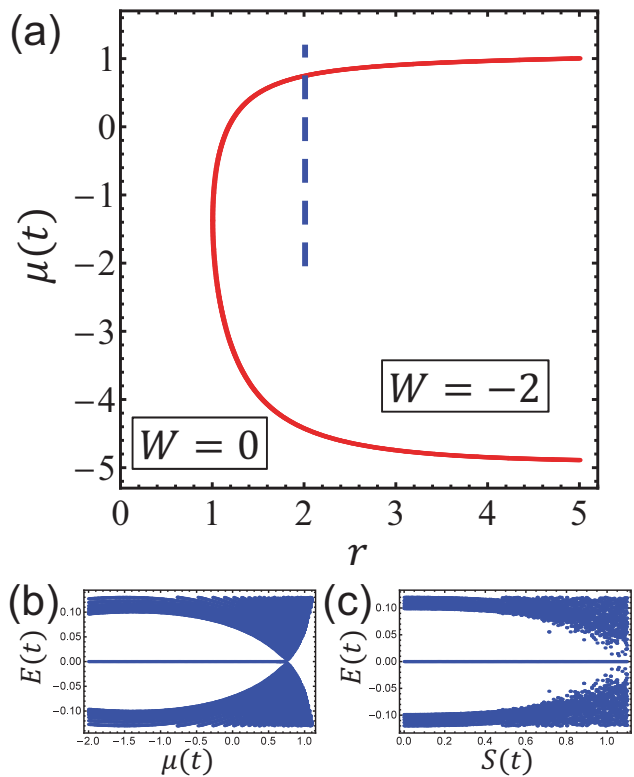}
\caption{\label{phaseAndSpectraCase3}
(a) Phase diagram of a 1D superconducting TI in the chemical potential-ratio of $\Delta_2$ to $\Delta_1$ ($\mu-r$) plane at $V_z=0$. Other parameters are the same as in Fig.~\ref{phaseAndSpectraCase1}(a). The corresponding winding numbers in different phase regions are indicated in the figure.
(b) Energy spectra $E$ vs chemical potential $\mu$ of a corresponding finite superconducting TI NW at $r=2$, i.e., along the dotted line in (a). The finite superconducting TI NW is modeled by a lattice of 800 sites.
(c) Energy spectra of the same system as in (b) calculated at different strengths of potential randomness $S$ at $\mu/t=-1$. 
}
\end{figure}

Figure~\ref{phaseAndSpectraCase3}(a) shows the phase diagram of an infinite 1D superconducting TI NW at zero magnetic field in the plane of the chemical potential ($\mu$) and the ratio ($r$) of $\Delta_2$ to $\Delta_1$, i.e., the $\mu-r$ plane. Several features can be found. One is that the phase boundaries defined by gap closing at $q=0$ and $\pi$ in the Brillouin zone and the phases of $|W|=1$ are disappeared, leaving only the third kind of boundaries and the $|W|=0$ and $2$ phases in the system. This is because without the Zeeman energy, the system at the special points of $q=0$ and $\pi$ in the Brillouin zone is always gapped. Another feature is that a prerequisite for the presence of a nontrivial topological phase is $r>1$, namely $\Delta_1<\Delta_2$.

Figure ~\ref{phaseAndSpectraCase3}(b) shows the energy spectra of a corresponding finite superconducting TI NW at $r=2$. It is seen that two pairs of zero-energy MF states are present in the $|W|=2$ phase, which are the Kramers pairs of MFs as the presence of time reversal symmetry at zero magnetic field. In order to test the stability of the two pairs of MF states, we have performed the calculations for the energy spectra of the system in a $|W|=2$ phase in the presence of disorders. Figure~\ref{phaseAndSpectraCase3}(c) shows the results of the calculations. Here, in the calculations, the disorders are modeled by random site potential fluctuations according to the
Gaussian distribution $\langle U(n_{i})U(n_{j})\rangle=S^{2}\delta_{ij}$
with a average value of $\overline{U}=0$ but different values of variance $S$. It can be found that as the strength $S$ increases, the gap will be filled
with normal quasiparticle states. However, the Kramers pairs of MFs are still protected
by a energy gap unless the disorder becomes too strong, demonstrating
the robustness of the Kramers pairs of MFs.

\section{Signature of MFs in Josephson junction devices}

We have by now obtained much information about the phase diagrams and the energy spectra of single TI NWs. However, there are two important questions which still remain to be addressed. (1) How do we detect the pair number of MFs at each phase?
(2) Is there any observable effect arising from the sign $"\pm"$ of the phase label?

Obviously, the answer to these questions can not come out from the measurements of a device made from a single segment of a TI NW and we
have to consider constructing
a Josephson junction device \cite{prl103.107002,prb87.104513} containing two segments of superconducting TI NWs.
Let us recall the connection between the MFs and the observables.
A generic signature of MFs in transport measurements of a single nanowire
is zero-bias conductance peaks (ZBPs). However, in our case, it cannot
tell us the number of MFs since multiple MFs can peacefully locate
at the same end of the nanowire at zero energy. While in a Josephson junction
device made from two segments of superconducting TI nanowires, the MFs near the junction are hybridized into Andreev bound
states,\cite{sr4.7261} which will be captured in the tunnel spectroscopy.
Therefore, the detection of the ZBPs and the peaks of Andreev bound states in such a Josephson junction device would eliminate most of other mechanisms and manifest
uniquely the existence of MFs.

\begin{figure}
\centering\includegraphics[scale=0.6]{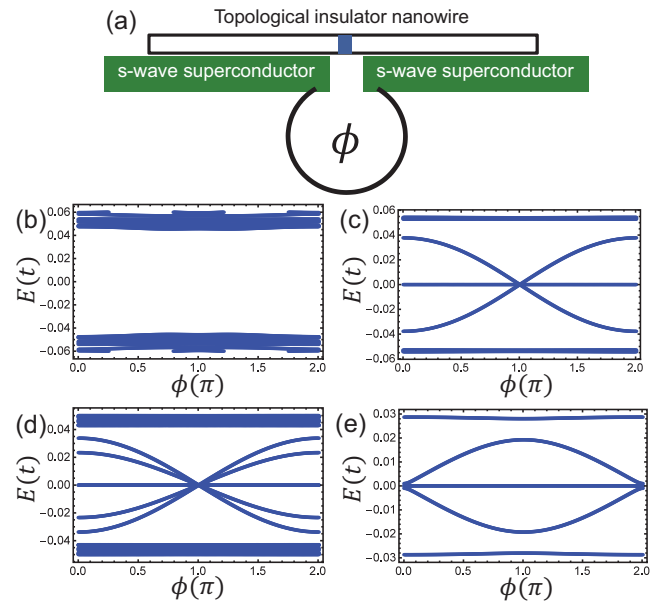}\caption{\label{JosephsonJunction}(a) Schematic for a TI NW Josephson junction device.
Here the phase difference $\phi$ between the two s-wave superconductors is assumed to be tuned through the flux in a SQUID configuration. (b) to (e) Evolutions of the energy spectra as a function of phase difference $\phi$ for a device built from superconducting TI NWs as considered in Fig.~\ref{phaseAndSpectraCase1}(a), i.e., in the case of $\Delta_1>\Delta_2$, with $V_{z}/t=0.5$ and $\mu/t=-2$ ($W=0$) in (b), $V_{z}/t=1.5$ and $\mu/t=0$ ($W=-1$) in (c), $V_{z}/t=1$ and $\mu/t=-1$ ($W=-2$) in (d), and $V_{z}/t=3.2$ and $\mu/t=-3.2$ ($W=-1$) in the left nanowire and $V_{z}/t=3.67$ and $\mu/t=-3.2$ ($W=1$) in the right nanowire in (e). Note that in the calculations for (b) to (e), each nanowire is modeled by a lattice of 600 sites and the two nanowires in each device is connected by a weak link modeled by assuming a smaller hopping parameter of $0.5t$ between the two connecting sites. Other unspecified parameters are the same as in Fig.~\ref{phaseAndSpectraCase1}(a).}
\end{figure}

We confirm the above scenario by proposing a SQUID setup as in
Fig.~\ref{JosephsonJunction}(a) and calculating the energy spectra as
a function of the phase difference $\phi(=\theta_{R}-\theta_{L})$ between the two superconducting NW segments in the device. Here, we note that the device consists of a weak link in the middle modeled by assuming a smaller hopping parameter between the connecting sites of the superconducting NW segments and the phase
difference $\phi$ between the
two segments can be tuned through the flux of the SQUID.
Figures~\ref{JosephsonJunction}(b)-\ref{JosephsonJunction}(e) show the energy spectra of the SQUID device at different values of $\phi$ for the superconducting NW segments at different topological phases.
In a trivial phase [Fig. \ref{JosephsonJunction}(b)],
there is no persistent zero energy states.
On the contrary, in a case when the superconducting NW segments are in a topological phase
[see Figs.~\ref{JosephsonJunction}(c)-\ref{JosephsonJunction}(e)], zero energy states are present
for all values of $\phi$ and the number of the subgap state crossing points
is odd (here just one crossing point) in one phase period region, which inherit the general feature of hybridized MFs.\cite{prl105.077001}
We can further observe that the pair number
of Andreev bound states is exactly the pair number of MFs, see Fig.~\ref{JosephsonJunction}(c) for an example of one pair and Fig.~\ref{JosephsonJunction}(d) for an example of two pairs. Thus, the measurements of the number of subgap Andreev bound states
in the SQUID device would provide information about the winding numbers or the pair number of MFs in each superconducting TI NW segment.\cite{prb87.104509}

If we further assume that $\mu$ and $V_{z}$ of the two superconducting segments are separately tunable,
as in the experiment setup proposed in Ref.~\onlinecite{prl108.067001}, we can
make the case that the two superconducting TI NW segments are in different topological phases. Figure~\ref{JosephsonJunction}(e) shows the energy spectra of the SQUID device in such a case in which the MF from the left NW segment in the $W=1$ phase interacts with the MF from the right NW segment in the
$W=-1$ phase. It is seen that the cross point of the topological protected subgap Andreev bound states is shifted from $\phi=\pi$ to $\phi=0$,
which indicates the appearance of sign $"-"$ in the phase index $-W$ of one NW segment as compared with the phase index $W$ of the other NW segment. As a consequence,
an intrinsic phase difference of $\pi$ is present between the two MF states and the formation of a topological
$\pi$ Josephson junction is realized.\cite{prb87.100506}

\section{Conclusions}

The phase diagrams and energy spectra of MF bound states in 1D TI NWs are studied. The chiral symmetry as well as the
particle-hole symmetry make the MF pair number in the systems labelled by a topological
invariant winding number $W$. In the presence of the superconducting pairing
potentials between both the same and different orbitals, $W$ could take
values from $-2$ to $2$. The energy spectra of MFs in a single nanowire are calculated and analyzed and the effect
of multiple MFs in a Josephson junction device is examined.
It is proposed that the pair number of MFs can be extracted in transport measurements of a Josephson junction device,
as it can be mapped to the number of pairs of topological protected subgap bound states in the device.
The effect of the sign in the winding numbers in a Josephson junction device has also been discussed
and sign extraction procedure has been proposed. The multiple MFs are non-Abelian anyons as in the case of single isolated MFs, which could have potential applications in topological quantum computation.\cite{prx4.021018,prl113.246401,prb93.045417} This study shows that the operation of both single and multiple pairs of MFs can
be achieved in a TI NW device.

\section*{ACKNOWLEDGMENTS}

The authors are grateful to Martin Leijnse for stimulating discussions.
This work was supported by the Ministry of Science and Technology of China (MOST) through the National Key Basic Research Program of China
(Grants No.~2012CB932703, No.~2012CB932700, and 2016YFA0300601), the National Natural Science Foundation of China (Grants No.~91221202, No.~91421303, No.~61321001, and No.~11604005), and the Swedish Research Council (VR). GYH would also like to acknowledges financial support from the China Postdoctoral Science Foundation (Grant No.~2016M591001).

\end{document}